# Nucleon and pion electromagnetic structure and constituent quark form factors

F. Cardarelli [1], E. Pace [1,2], G. Salmè [3], S. Simula [3]

[1] Istituto Nazionale di Fisica Nucleare, Sezione Tor Vergata, Via della Ricerca Scientifica 1, I-00133 Roma, Italy

[2] Dipartimento di Fisica, Università di Roma "Tor Vergata", Via della Ricerca Scientifica 1, I-00133 Roma, Italy

[3] Istituto Nazionale di Fisica Nucleare, Sezione Sanità, Viale Regina Elena 299, I-00161 Roma, Italy

**Abstract.** We have evaluated nucleon and pion electromagnetic form factors using for the first time eigenfunctions of a light-front mass operator, reproducing a large set of energy levels, and constituent quark form factors in the one-body current. The comparison with the experimental data yields valuable information on the electromagnetic structure of light constituent quarks and on the different role played by their Dirac and Pauli form factors.

The investigation of the electromagnetic (e.m.) form factors of hadrons, within relativistic constituent quark (CQ) models, has been widely [1, 2, 3, 4] pursued for extending the range of applicability of the CQ picture of hadrons. Our approach [3, 4] is based on the CQ model embedded in the light-front (LF) formalism, which has many interesting features [5]. Among the properties relevant for what follows, there are: i) the spacelike e.m. form factors, evaluated in a frame where the plus component of the momentum transfer $q_+ = q_0 + q_z = 0$, are not affected by the contribution of the pair creation from the vacuum [6], and ii) the centre of mass motion in the wave function of the interacting system can be exactly separated. As to the latter, the LF hamiltonian of the interacting system, $P_- = \frac{\mathcal{M}^2 + P_\perp^2}{P_+}$, allows a straightforward separation of the coordinates of the centre of mass, $(P_\mp = P_0 \mp P_z,\ P_\perp \equiv [P_x, P_y])$, from the intrinsic ones contained in the mass operator defined by, e.g., $\mathcal{M} = M_0 + \mathcal{V}$, with $M_0$ the free mass and $\mathcal{V}$ a Poincaré invariant interaction.

The main features of our approach are: i) the hadron state is eigenvector of a LF mass operator, constructed from the effective $q\bar{q}$ and $qq$ interaction of [7], that reproduces a lot of energy levels; ii) the configuration mixing, due to the one-gluon-exchange (OGE) part of the effective interaction, leads to high momentum components and SU(6) breaking terms in the hadron wave function; iii) Dirac and Pauli form factors for the CQ's, as suggested by their quasi-particle nature (cf. [8]), are introduced in the one-body e.m. current.

The LF hadron wave functions are eigenvectors of the mass operator and of the non-interacting angular momentum operators $j^2$ and $j_z$. Disregarding the color degree of freedom, the LF wave function is (see [1, 4])

$$\langle\{\xi_i q_{i\perp};\nu_i\tau_i\}|\Psi_{\mathcal{H}}\rangle \;=\; \mathcal{N}_{\mathcal{H}} \sum_{\{\nu_i'\}} \langle\{\nu_i\}|\mathcal{R}^\dagger|\{\nu_i'\}\rangle\langle\{q_i;\nu_i'\tau_i\}|\psi_{\mathcal{H}}\rangle \qquad (1)$$

where the curly braces $\{\ \}$ indicates $i = 1, n_q$ with $n_q = 2(3)$ for the mesons (baryons); $\xi_i$, $q_{i\perp}$ and $q_{iz}$ are intrinsic LF variables; $\mathcal{N}_{\mathcal{H}}$ is the proper normalization factor; $\nu_i'(\tau_i)$ is the third component of the quark spin (isospin); $\mathcal{R}^\dagger = \left[\prod_{j=1}^{n_q} R_M^\dagger(q_{j\perp},\xi_j,m_j)\right]$is a unitary operator, with $R_M(q_{j\perp},\xi_j,m_j)$ being the generalized Melosh rotation. The equal-time state $|\psi_{\mathcal{H}}\rangle$ is an eigenvector of the transformed mass operator $M = \mathcal{R}\mathcal{M}\mathcal{R}^\dagger = \mathcal{R}M_0\mathcal{R}^\dagger + \mathcal{R}\mathcal{V}\mathcal{R}^\dagger$, that becomes $M = M_0 + V$. The interaction $V = \mathcal{R}\mathcal{V}\mathcal{R}^\dagger$ must fulfil the translational and rotational invariances, and be independent of the total momentum (cf. [5]). Since it turns out that $M_0 = \sum_{i=1}^{n_q}\sqrt{m_i^2 + q_i^2}$, we have chosen [3, 4] $M$ equal to the effective hamiltonian proposed for mesons and baryons in [7]. Thus, our equal-time hadron state $|\psi_{\mathcal{H}}\rangle$ is eigenvector of the following mass operator

$$H_{\mathcal{H}}\;|\psi_{\mathcal{H}}\rangle \;\equiv\; \left[\sum_{i=1}^{n_q}\sqrt{m_i^2+q_i^2} + \sum_{i\neq j=1}^{n_q} V_{ij}\right]\;|\psi_{\mathcal{H}}\rangle \;=\; M_{\mathcal{H}}\;|\psi_{\mathcal{H}}\rangle \qquad (2)$$

where $M_{\mathcal{H}}$ is the mass of the hadron, and $V_{ij}$ the effective $q\bar{q}$ or $qq$ potential for mesons or baryons, respectively. The interaction is composed by a OGE term (dominant at short separations) and a linear confining term (dominant at large separations). As in [7], the values $m_u = m_d = 0.220\ GeV$ have been adopted. The wave equation (2) has been solved by expanding the wave function onto a (truncated) harmonic oscillator basis and then applying the Rayleigh-Ritz variational principle (see [4]). It should be pointed out that the relativistic mass operator (2) reproduces a large set of hadron energy levels [7] and generates a huge amount of configuration mixing, due to the presence of the OGE part of the interaction, that also determines the hyperfine mass-splitting.

In [3, 4] we have evaluated the nucleon and pion form factors using the eigenstates of Eq.(2), whereas in the current literature only the effects of the confinement scale are considered, through gaussian or power-law wave functions (cf. for the nucleon [1, 2]), without taking care of the mass eigenvalues. In particular we have shown that, assuming pointlike CQ's, the calculated nucleon and pion form factors sharply differfrom the experimental data, because of configuration mixing effects. Therefore the possibility of a non trivial e.m.

structure for extended CQ's, summarizing the underlying degrees of freedom, has been considered, adopting the following one-body component of the hadron e.m. current

$$\mathcal{I}^{\nu}_{\mathcal{H}} = \sum_{j=1}^{n_q} I^{\nu}_j = \sum_{j=1}^{n_q} \left( e_j \gamma^{\nu} f^j_1(Q^2) \; + \; i\kappa_j \frac{\sigma^{\nu}{}_{\mu} q^{\mu}}{2m_j} f^j_2(Q^2) \right) \tag{3}$$

where $e_j$ is the charge of the j-th quark, $\kappa_j$ the corresponding anomalous magnetic moment, $f^j_{1(2)}$ its Dirac (Pauli) form factor, and $Q^2 \equiv -q \cdot q$ the squared four-momentum transfer. The values of $\kappa_u$ and $\kappa_d$ can be fixed by the request of reproducing the experimental nucleon magnetic moments, obtaining $\kappa_{u(d)} = 0.085$ $(-0.153)$ [4]. In spite of their smallness such values affect the nucleon anomalous magnetic moments by $\sim 40\%$. The CQ Dirac and Pauli form factors have been parametrized in a simple way (the Dirac form factor = monopole + dipole, and the Pauli one = dipole + octupole); the values of the parameters have been estimated through a standard minimization procedure using the experimental nucleon and pion form factors in a wide range of momentum transfer. The results of [4] are shown in Figs. 1 and 2 (solid lines). It can be seen that our calculations, though obtained in the framework of the one-body approximation for the hadron e.m. current (i.e. by disregarding the two-body currents necessary for fulfilling both the gauge and rotational invariances, see [5]), are in nice agreement with the data. The CQ form factors, appearing in the effective one-body current Eq. (3), are presented in Fig. 3.

As previously mentioned, the underlying degrees of freedom frozen in the extended CQ influence their properties, such as the form factors and the anomalous magnetic moments. In order to undertake a more detailed investigation, in this contribution we would focus on the different role played by the Dirac and Pauli form factors of the CQ's. To this end the comparison with the experimental data is carried out in Figs. 1 and 2, considering both the full calculation and the one performed by retaining the CQ Dirac form factor only, i.e. assuming $\kappa_{u(d)} = 0$ (dashed lines). The following comments are in order: i) $G^p_E$ is strongly affected by $f^q_2$ for $Q^2 > 1(GeV/c)^2$; ii) $G^n_E$ changes its sign when $f^q_2$ is switched off; iii) the nucleon magnetic form factors are very sensitive at low $Q^2$ to the CQ Pauli form factor and dominated by the Dirac one at high $Q^2$; iv) the pion form factor is only slightly influenced by $f^q_2$, according to [3]. It is worth noting that the CQ form factors, shown in Fig.3, yield a quark radius, defined as $< r^{u(d)}_1 >^2 = -6 \; df^{u(d)}_1(Q^2)/dQ^2$ at $Q^2 = 0$, equal to $< r^{u(d)}_1 > = 0.51$ $(0.42) fm$; such values are similar to the ones obtained in Refs. [8, 9] and in our exploratory analysis of the pion data [3] (cf. also the dotted line in Fig. 3).

To sum up, within the LF formalism, we have given a simultaneous description of the pion and nucleon form factors adopting for the first time the eigenfunctions of a LF mass operator, with correct mass eigenvalues, and a one-body approximation for the e.m. current, containing CQ form factors. The same model has been applied in a parameter-free calculation of the magnetic form factor of the $N - \Delta$ transition, which has been recently completed [10] achieving a very satisfactory result. The improvement of our approach including the contribution of two-body e.m. currents is in progress.

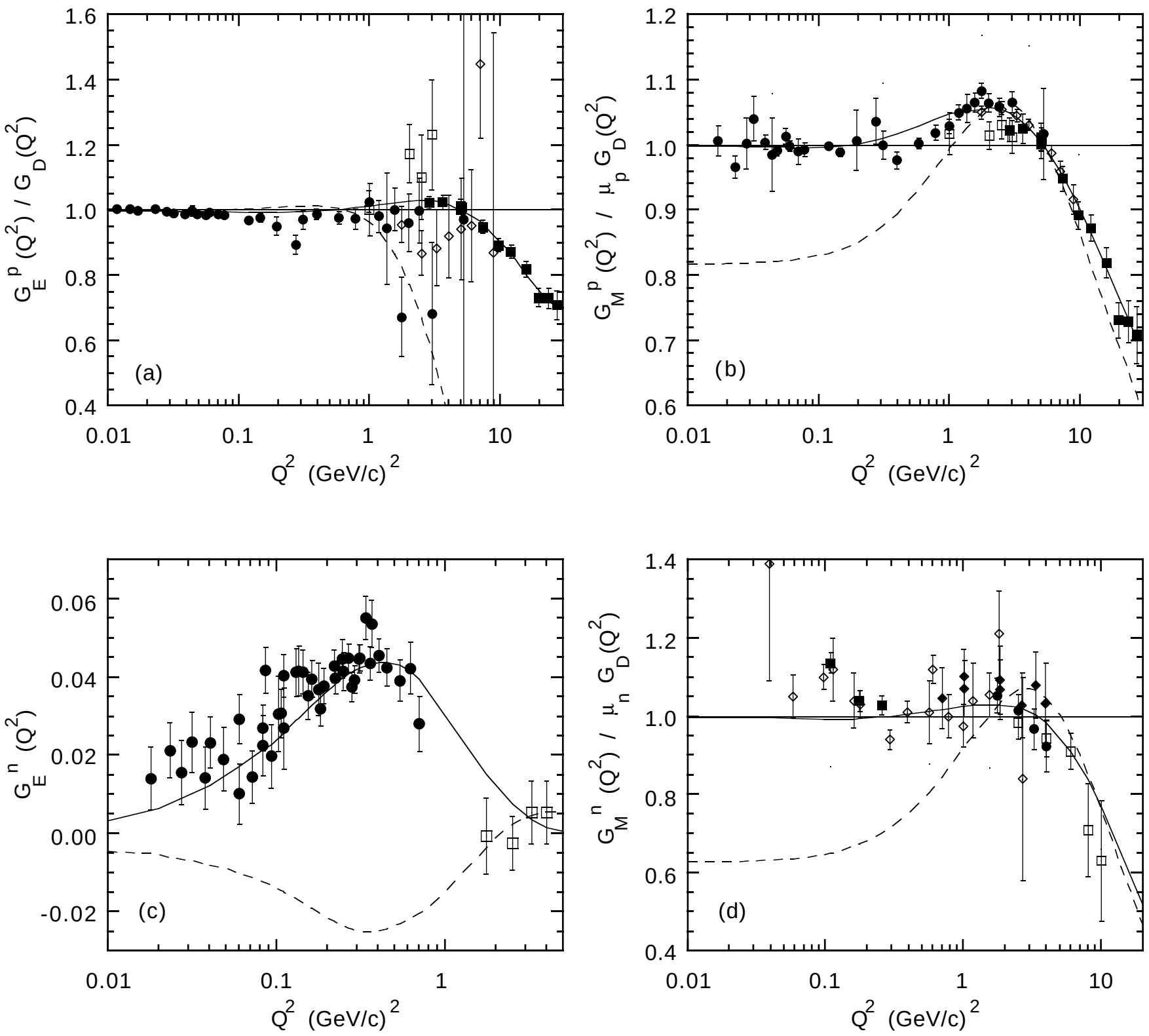

**Figure 1.** - (a) The proton form factor $G_E^p/G_D$ vs. $Q^2$. Solid line: $G_E^p/G_D$ obtained in [4] by using the nucleon wave function corresponding to the interaction of [7] and the nucleon e.m. current with CQ form factors. Dashed line: the same as the solid line, but with $\kappa_{u(d)} = 0$. Experimental data are quoted in detail in [4]. - (b) The same as in Fig. 1a, but for $G_M^p/(\mu_p\ G_D)$.- (c) The same as in Fig. 1a, but for $G_E^n$. - (d) The same as in Fig. 1a, but for $G_M^n/(\mu_n\ G_D)$.

## References

1. P.L. Chung and F. Coester: Phys. Rev. **D44**, 229 (1991); S. Capstick and B. Keister: Phys. Rev. **D51**, 3598 (1995).

2. Z. Dziembowski : Phys. Rev. **D37**, 778 (1988); I.G. Aznaurian: Phys. Lett. **B316**, 391 (1993); H. J. Weber: Phys. Rev. **D49**, 3160 (1994); F. Schlumpf: Jou. of Phys. **G20**, 237 (1994).

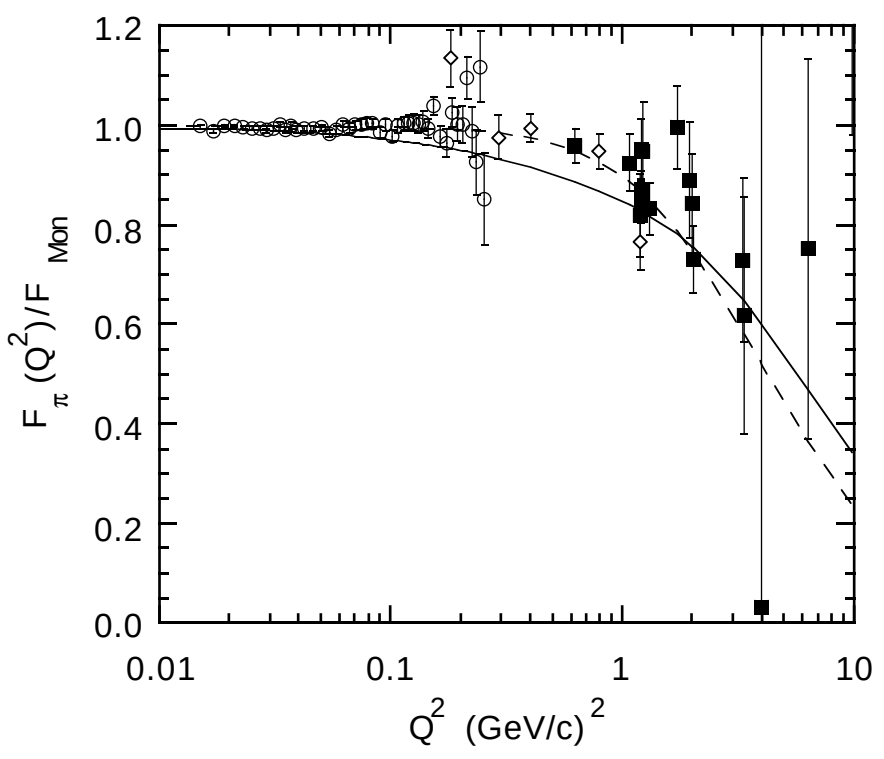

**Figure 2.** The charge form factor of the pion $F_\pi(Q^2)/F_{Mon}(Q^2)$ vs. $Q^2$, with $F_{Mon}(Q^2) = 1/(1 + Q^2/0.54)$. The solid line has been obtained in [4] by using the pion wave function corresponding to the interaction of [7], and CQ form factors. Dashed line: the same as the solid line, but with $\kappa_u = \kappa_d = 0$. Experimental data are quoted in detail in [4].

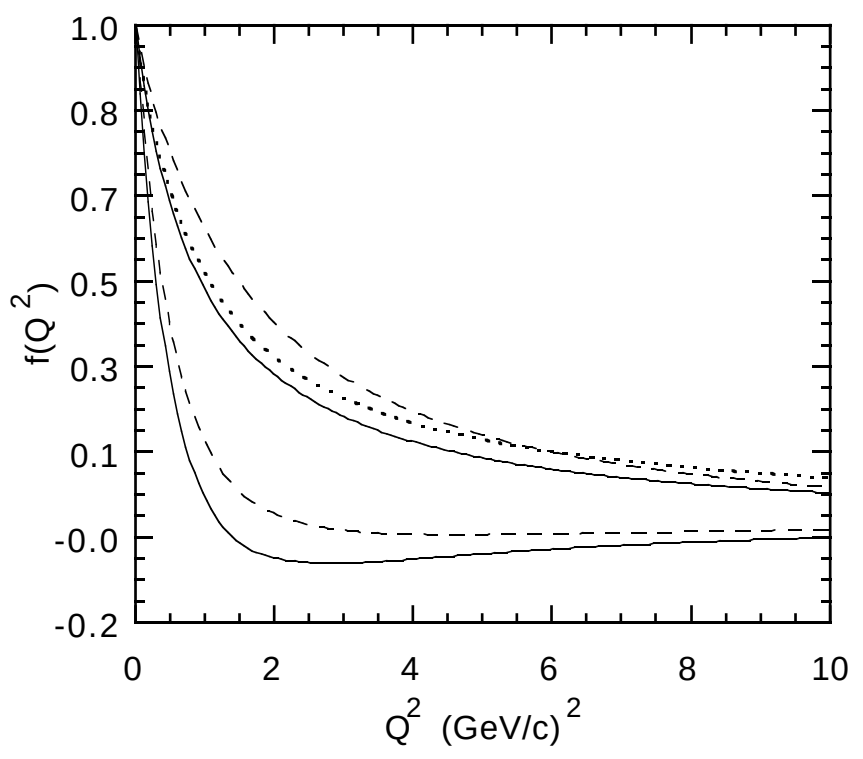

**Figure 3.** The constituent quark form factors, extracted from the analysis of the nucleon and pion form factors (see Figs. 1 and 2), vs. $Q^2$. Solid lines represent $f_1^u(Q^2)$ (upper) and $f_2^u(Q^2)$ (lower), respectively; dashed lines represent $f_1^d(Q^2)$ (upper) and $f_2^d(Q^2)$ (lower), respectively. For comparison $f_1^u(Q^2) = f_1^d(Q^2)$ obtained in [3] is also shown (dotted line). (After [4])

3. F. Cardarelli, I.L. Grach, I.M. Narodetskii, E. Pace, G. Salmè and S. Simula: Phys. Lett. **B332**, 1 (1994); nucl-th 9507038, submitted to Phys. Rev. D, brief report. F. Cardarelli, I.L. Grach, I.M. Narodetskii, G. Salmè and S. Simula: Phys. Lett. **B349**, 393 (1995); Phys. Lett. **B359**, 1 (1995).
4. F. Cardarelli, E. Pace, G. Salmè and S. Simula: Phys. Lett. **B357**, 267 (1995); F. Cardarelli and S. Simula: to be published.
5. For a review, see B.D. Keister and W.N. Polyzou: Adv. in Nucl. Phys. **20**, 225 (1991) and F. Coester: Progress in Part. and Nucl. Phys. **29**, 1 (1992).
6. See for instance M. Sawicki: Phys. Rev. **D46**, 474 (1992).
7. a) S. Godfrey and N. Isgur: Phys. Rev. **D32**, 185 (1985); b) S. Capstick and N. Isgur: Phys. Rev. **D34**, 2809 (1986).
8. U. Vogl, M. Lutz, S. Klimt and W. Weise: Nucl. Phys. **A516**, 469 (1990).
9. B. Povh and J. Hufner: Phys. Lett. **B245** (1990) 653.
10. F. Cardarelli, E. Pace, G. Salmè and S. Simula: preprint INFN-ISS 95/12 (nucl-th 9509033), submitted to Phys. Lett. B.